\documentclass{elsart}
\usepackage{graphicx}

\begin{document}
\begin{frontmatter}

\title{The effect of non-ideal market conditions on option pricing}
\author{Josep Perell\'o and Jaume Masoliver}
\address{Departament de F\'{\i}sica Fonamental, 
Universitat de Barcelona, Diagonal, 647, 08028-Barcelona, Spain}
\date{\today}
\maketitle

\begin{abstract}
Option pricing is mainly based on ideal market conditions which are well represented by the Geometric Brownian Motion (GBM) as market model. We study the effect of non-ideal market conditions on the price of the option. We focus our attention on two crucial aspects appearing in real markets: The influence of heavy tails and the effect of colored noise. We will see that both effects have opposite consequences on option pricing.  
\end{abstract}
\end{frontmatter}

\section{Introduction}

Option pricing lies, from the very beginning, at the heart of mathematical finance as scientific discipline. In 1900, Bachelier proposed the arithmetic Brownian motion for the dynamical evolution of stock prices as a first step towards obtaining a price for options \cite{cootner}. Since then and specially after the landmark work of Black and Scholes \cite{BS} and Merton \cite{mertonBS}, the underlying asset of any option is usually assumed to be driven by Brownian motion, being the most ubiquitous model the geometric Brownian motion (GBM):
\begin{equation}
\frac{\dot S}{S}=\mu+\frac 12\sigma^2+\sigma\xi(t),
\label{gbm}
\end{equation}
where $S(t)$ is the price of stock (or the value of an index) at time $t$, $\mu+\sigma^2/2$ is the drift, $\sigma$ is the volatility and $\xi(t)$ is Gaussian white noise with zero mean and correlation  $\langle\xi(t)\xi(t')\rangle=\delta(t-t')$. Associated with stock price $S(t)$ there is another random process: the return. In continuos-time finance this is defined by $R(t)=\ln S(t)/S(t_0)$. For the GBM 
(\ref{gbm}) one can see, after using It\^o lemma \cite{perello1}, that the return obeys the following SDE
\begin{equation}
\dot{R}=\mu+\sigma\xi(t),
\label{gbmR}
\end{equation}
from which we see that $\mu$ is the mean return rate. Since, by definition, the initial return is zero, then in terms of the integral of $\xi(t)$ which is taken in the It\^o sense, we can explicitly write the return as
\begin{equation}
R(t)=\mu(t-t_0)+\sigma\int_{t_0}^t\xi(t')dt',
\label{gbmRsol}
\end{equation}
where the integral of $\xi(t)$ appearing in the rhs of this equation is simply the Wiener process $W(t)$. Note that, independently of the stock price $S(t_0)$, the return at initial time $t_0$ is always zero since $R(t_0)=\ln S(t_0)/S(t_0)=\ln 1=0$.

However, the market model given by Eq.~(\ref{gbm}) or Eqs.~(\ref{gbmR})--(\ref{gbmRsol}) does not fully account for all properties observed in real markets. Some of them are quite crucial such as the existence of heavy tails and negative skewness in probability distributions, evidence of self-scaling (at least for high frequency data) \cite{mante95}, leverage \cite{bouch2001} and even the existence of mild correlations in the increments of return \cite{lo}. All of this, undoubtedly has to affect option pricing. Therefore, the theoretical price for the option obtained through the assumption that markets are governed by the GBM would not be adjusted to reality. Our main objective here is to elucidate how some of the above properties of real markets may affect the price of options. Specifically we will obtain an option price for market models that present fat tails and self-scaling. After that, we will relax the so called efficient market hypothesis \cite{fama} and assume the existence of auto correlations in stock prices. We restrict the analysis to European call options, although it can be easily generalized to the European put and other options with fixed exercise date. The extension to American options seems to be more involved.

The starting point of this work is the {\it martingale option price}. As was shown in early eighties, the Black-Scholes (B-S) option price can also be obtained using martingale methods \cite{harrison}. This is a shorter, although more abstract way, to derive an expression for the call price. The main advantage is that one only needs to know the probability density function governing market evolution. The main inconvenience is that there is no known hedging strategy as in the case of B-S theory. 

The probability distribution of some of the more realistic market models we will use is only known through its characteristic function. We will therefore derive a formula which will allow us to obtain, via martingale price, the option price in terms of the characteristic function of the market model chosen. Having the necessary tools to address the problem, we will apply them to a recently presented market model which shows self-similarity, fat tails and finite moments \cite{mmp}. We will obtain an  option price this model and compare it with B-S price.

The second problem we want to address is the effect of colored noise in the price of options. The GBM model supposes that prices are driven by Gaussian white noise, this implies that prices $S(t)$, at different times, are uncorrelated which, for certain times scales, does not seem to be completely adjusted to reality. We will thus assume that prices are driven by colored noise and obtain the corresponding fair price for the European call. 

The paper is organized as follows, after this introduction, in Sect.~\ref{mart} we review, from a physicist's point of view, the martingale price for the European call. In Section~\ref{four} we obtain the call price formula using Fourier analysis and apply it in Sect.~\ref{heav} to heavy tail and self-similar processes. In Sect.~\ref{cor} we study the effect of colored noise on option pricing. Conclusions are drawn in Sect.~\ref{conc}.

\section{The martingale option price\label{mart}}

We start by recalling that an European call option is a derivative giving to its owner the right but not the obligation to buy a share at a fixed date $T$ (the maturity) for a certain price $K$ (the strike or striking price). The call option contract is basically specified by its gain at maturity. If $S=S(T)$ is the share price at time $T$ this gain, called payoff, is given by
\begin{equation}
(S-K)^+\equiv{\rm max}[S-K,0].
\label{payoff}
\end{equation}

As was shown in the early eighties \cite{harrison}, the B-S option price can also be obtained using martingale methods. This is a shorter, although more abstract way, to derive an expression for the call price. The main advantage being that one only needs to know the probability density function governing market evolution. 

Before proceeding further let us briefly explain, in layman terms and without pretending to be rigorous, what a martingale is. Suppose $X(t)$ is a well defined random process such that its previous history is perfectly known. That is, we keep record, and hence know of, all values taken by $X(t)$ before time $t$. Under these conditions the process is a martingale if the conditional expected value 
$\langle X(t)| X(t_0)=x_0\rangle$ for any $t>t_0$ is given by
\begin{equation}
\langle X(t)| X(t_0)=x_0\rangle=x_0,\qquad(t>t_0).
\label{martingale}
\end{equation}
An example. Under quite general conditions any driftless process driven by zero-mean noise is a martingale \cite{baxter}. If $X(t)$ follows the SDE: $\dot{X}=F(t)$ where $F(t)$ is a zero-mean random process, then
$$
X(t)=x_0+\int_{t_0}^{t}F(t')dt',
$$
where $x_0=X(t_0)$ is known and $F(t)$ is supposed to be integrable, at least in the sense of generalized functions. Since $\langle F(t)\rangle=0$, we have $\langle X(t)|x_0\rangle=x_0$. Hence $X(t)$ is a martingale. 

In what follows we generalize the GBM model (\ref{gbm})--(\ref{gbmRsol}) to include driving noises other than the Wiener process and thus trying to account for some features of real markets such as fat tails, self-scaling and, eventually, correlations. We therefore proposed as market model the one whose return is given by (see Eq. (\ref{gbmRsol}))
\begin{equation}
R(t)=\mu(t-t_0)+\int_{t_0}^tF(t')dt',
\label{newmodel0}
\end{equation}
where $\mu$ is the (constant) mean return rate, $F(t)$ is zero-mean noise and, if needed, the integral of $F(t)$ should be interpreted in the sense of It\^o. We define the {\it zero-mean return} $X(t)$ by
\begin{equation}
X(t)=\int_{t_0}^tF(t')dt'.
\label{X}
\end{equation}
In generalizing the GBM for option pricing we have to assume that the zero-mean return $X(t)$ does not allow for arbitrage opportunities. In other words, it must avoid money profits for free without taking any risk. To this end, as was proved by Harrison {\it et al.} \cite{harrison2}, it suffices that $X(t)$ be a noise of unbounded variation. We will assume this condition in what follows. 

One can easily show from Eq. (\ref{X}) that 
$$
\langle X(t_2)|X(t_1)\rangle=X(t_1)
$$ 
for any $t_2>t_1$. Hence, the zero-mean return is a martingale. Moreover, in terms of $X(t)$ the return is 
\begin{equation}
R(t)=\mu(t-t_0)+X(t),
\label{newmodel}
\end{equation}
in consequence, $R(t)$ obeys the following SDE 
\begin{equation}
\dot{R}(t)=\mu+F(t),
\label{newmodeldif}
\end{equation}
which is the generalization of SDE (\ref{gbmR}).

We now return to options. The equivalent martingale measure theory imposes the condition that, in a ``risk-neutral world", the stock price $S(t)$ evolves, on average, as a riskless security and discards any underlying process that allows for arbitrage opportunities. Thus if at time $t$ we buy a European call on stock having price $S$, then in terms of the risk-neutral density $p^*(R,t|t_0)$ it is possible to express the price for the European call option by defining its value as the discounted expected gain due to holding the call. That is \cite{harrison}
\begin{equation}
C(S,t)=e^{-r(T-t)}\int_{\ln(K/S)}^{\infty}\left(Se^R-K\right)p^*(R,T|t)dR,
\label{call1}
\end{equation}
where $T$ and $K$ are the maturity and strike of the option. 

Let us define the risk-neutral density $p^*(R,t|t_0)$. This density, which  is also called the ``equivalent martingale measure'' associated with return $R(t)$ under ``risk-neutrality'', is the probability density function (pdf) of a process $R^*(t)$ which we call {\it risk-neutral return} and such that 
\begin{equation}
\left\langle e^{R^*(t)}|t_0\right\rangle=e^{r(t-t_0)},
\label{risknuetralcond}
\end{equation}
where $r$ is the risk-free interest rate ratio. Since the stock price associated with $R^*$ is given by $S^*(t)=S(t_0)e^{R^*(t)}$, condition (\ref{risknuetralcond}) imposes that risk-neutral prices $S^*$ must grow on average as a riskless security.

Note that we can easily prove that stock price $S$ is always greater than call price. In effect, from (\ref{call1}) we have
$$
C(S,t)\leq S e^{-r(T-t)}\int_{-\infty}^{\infty} e^R p^*(R,T|t)dR=
S e^{-r(T-t)}\left\langle e^{R^*(T)}|t\right\rangle
$$
and, after using Eq. (\ref{risknuetralcond}) we conclude that 
\begin{equation}
C(S,t)\leq S.
\label{boundS}
\end{equation}
We now assume that the market model for the evolution of the underlying stock is given by Eq. (\ref{newmodeldif}). Then, analogously to Eq. (\ref{newmodel}) for the return $R(t)$, we impose that the risk-neutral return $R^*(t)$ be written as
\begin{equation}
R^*(t)=m^*(t-t_0)+X(t),
\label{riskneutralR}
\end{equation}
where $m^*(t)$ is a deterministic function and $X(t)$ is given by Eq. (\ref{X}). In order to find the expression for $m^*(t)$, observe that 
$$
\left\langle e^{R^*(t)}\right\rangle=
e^{m^*(t)}\left\langle e^{X(t)}\right\rangle,
$$ 
and from the risk neutrality condition (\ref{risknuetralcond}) we see that
$$
m^*(t-t_0)=r(t-t_0)-\ln\left\langle e^{X(t)}\right\rangle.
$$ 
The average $\langle e^{X(t)}\rangle$ is obtained from the known probability distribution of the driving noise $X(t)$. Indeed, let $\varphi_X(\omega,t|t_0)$ be the characteristic function of $X(t)$,
\begin{equation}
\varphi_X(\omega,t|t_0)\equiv
\left\langle e^{i\omega X(t)}|t_0\right\rangle,
\label{fcX1}
\end{equation}
then 
\begin{equation}
\langle e^{X(t)}\rangle=\varphi_X(-i,t|t_0)
\label{expX}
\end{equation}
and
\begin{equation}
m^*(t-t_0)=r(t-t_0)-\ln\varphi_X(-i,t|t_0).
\label{m*}
\end{equation}
For the GBM, the zero-mean return $X(t)$ is the Wiener process ({\it cf.} Eq. (\ref{gbmR})). Thus 
\begin{equation}
\varphi_X(\omega,t|t_0)=e^{-\sigma^2\omega^2(t-t_0)/2},
\label{fcXgbm}
\end{equation}
consequently $\ln\varphi_X(-i,t|t_0)=\sigma^2(t-t_0)/2$ is the spurious drift, and 
\begin{equation}
m^*(t-t_0)=(r-\sigma^2/2)(t-t_0).
\label{m*gbm}
\end{equation}

Finally, since $R^*(t)$ and $X(t)$ are related by the simple linear relation 
(\ref{riskneutralR}), the risk-neutral density $p^*$ can be written, in terms of the zero-mean return density $p_X$, as
\begin{equation}
p^*(R,t|t_0)=p_X(R-m^*(t-t_0),t|t_0),
\label{relationpdfs}
\end{equation}
and the price of the call is given by
\begin{equation}
C(S,t)=e^{-r(T-t)}
\int_{\ln(K/S)}^{\infty}\left(Se^R-K\right)p_X(R-m^*(T-t),T|t)dR.
\label{call2}
\end{equation}

\section{Martingale option price by Fourier analysis\label{four}}

As we have said, the main advantage of martingale pricing is that one can obtain a fair option price when the market obeys a random dynamics different than the geometric Brownian motion. In this case one only has to know the return distribution of the underlying and then, using Eq. (\ref{call2}), one readily gets a fair price for the option. Nevertheless, knowing an analytical expression of the pdf $p_X(x,t|t_0)$ may be, in practice, beyond our reach. There are however many situations in which we know the characteristic function of the model although one is not able to invert the Fourier transform and obtain the density $p_X(x,t|t_0)$. This is the case of non-Gaussian models such as L\'evy processes \cite{mante95} and some generalizations \cite{mmp}, the case of stochastic volatility models \cite{shobel,perello3} among others. For these cases, we will develop an option pricing based on a combination of martingale methods and harmonic analysis. 

We recall that the characteristic function (cf) of a random variable is the Fourier transform of its probability density function. Thus the characteristic function of the return will be given by:
$$
\varphi(\omega,t|t_0)=\left\langle e^{i\omega R(t)}|t_0\right\rangle =
\int_{-\infty}^{\infty} e^{i\omega x} p(R,t|t_0)dx.
$$
The starting point of our analysis is Eq. (\ref{call1}) that we write in the form
\begin{eqnarray}
C(S,t)=e^{-r(T-t)}
\left[\int_{-\infty}^{\infty}\Bigl(Se^{R}-K\Bigr)p^*(R,T|t)dR \right.
\qquad \qquad \nonumber \\
\left.-\int_{-\infty}^{\ln(K/S)}\Bigl(Se^{R}-K\Bigr)p^*(R,T|t)dR \right].
\label{C1}
\end{eqnarray}
The first integral on the right hand side (rhs) of this equation can be solved in closed form with the result (see Eq. (\ref{risknuetralcond}))
$$
\int_{-\infty}^{\infty}\Bigl(Se^{R}-K\Bigr)p^*(R,T|t)dR=
S\left\langle e^{R^*(T)}\left|\right.t\right\rangle-K=
Se^{r(T-t)}-K.
$$
As to the second integral on the rhs of Eq.~(\ref{C1}), we have
$$
I\equiv\int_{-\infty}^{\ln(K/S)}\Bigl(Se^{R}-K\Bigr)p^*(R,T|t)dR=
K\int_{0}^{\infty}\Bigl(e^{-z}-1\Bigl)p^*(-z-R_K,T|t)dR,
$$
where 
\begin{equation}
R_K\equiv \ln(S/K)
\label{moneyness}
\end{equation}
can be considered as the {\it return associated with moneyness}\footnote{The neologism ``moneyness" refers to the ratio $S/K$.} The inverse Fourier transform of the characteristic function
$$
p^*(R,T|t)=\frac{1}{2\pi}\int_{-\infty}^{\infty} e^{-i\omega R} \varphi^*(\omega,T|t) d\omega
$$
allows us to write this second integral as
\begin{equation}
I=\frac{K}{2\pi}\int_{-\infty}^{\infty}\varphi^*(\omega,T|t)e^{i\omega R_K}d\omega
\int_{0}^{\infty}e^{i\omega z}\Bigl(e^{-z}-1\Bigr)dz.
\label{I0}
\end {equation}
In the Appendix A we show that $I$ can be written as
\begin{equation}
I=-\frac{K}{2}+
\frac{K}{2\pi}\int_{-\infty}^{\infty}\varphi^*(\omega,T|t)\frac{e^{i\omega R_K}d\omega}{1-i\omega}
-i\int_{-\infty}^{\infty}\Bigl[e^{i\omega R_K}\varphi^*(\omega,T|t)-1\Bigr]
\frac{d\omega}{\omega}.
\label{I}
\end{equation}
Collecting results into Eq.~(\ref{C1}), we finally get
\begin{eqnarray}
C(S,t)=S-\frac{K}{2}e^{-r(T-t)}-\frac{K}{2\pi}e^{-r(T-t)}
\Biggl\{\int_{-\infty}^{\infty}\varphi^*(\omega,T|t)\frac{e^{i\omega R_K}d\omega}{1-i\omega}
\nonumber\\
-i\int_{0}^{\infty}\Bigl[e^{i\omega R_K}\varphi^*(\omega,T|t)-1\Bigr]
\frac{d\omega}{\omega}\Biggr\}.
\label{charcall2}
\end{eqnarray}
The representation~(\ref{charcall2}) is indeed very useful when the risk-neutral density $p^*$ is unknown but its characteristic function $\varphi^*$ is known. This would be indeed the case of more sophisticated market models such as those of next sections. A similar result has been given in \cite{scott} using a different but equivalent form of Eq.~(\ref{charcall2}) in order to numerically perform the integration knowing $\varphi^*(\omega,T|t)$. It is asserted in \cite{scott} that these Fourier methods allow a fast computing of the option price.

We close this section presenting an alternative price formula to that of Eq. (\ref{charcall2}) that is written in terms of the characteristic function of the zero-mean return instead of the risk-neutral cf $\varphi^*$. This price formula only involves real quantities and it can be more convenient in a number of cases. From Eq. (\ref{relationpdfs}) we easily see that 
\begin{equation}
\varphi^*(\omega,T|t)=e^{i\omega m^*(T-t)}\varphi_X(\omega,T|t),
\label{fcX}
\end{equation}
where $\varphi_X(\omega,T|t)$ is the characteristic function of the zero-mean return $X(t)$. Moreover, since $\langle X(t)|t_0\rangle=0$, the distribution of $X(t)$, $p_X(x,T|t)$, is symmetric around $x=0$. Then its Fourier transform is a real and even function of $\omega$ \cite{lukacs}, 
$$
\varphi_X(-\omega,T|t)=\varphi_X(\omega,T|t).
$$ 
These properties allow us to further simplify the integrals appearing in Eq.~(\ref{charcall2}). In effect, in terms of $\varphi_X(\omega,T|t)$ given by Eq.~(\ref{fcX}), the first integral on the rhs of Eq.~(\ref{charcall2}) (we call it $I_1$) reads
$$
I_1=
2\int_{0}^{\infty}\varphi_X(\omega,T|t)\frac{\cos\omega\alpha(T-t)-
\omega\sin\omega\alpha(T-t)}{1+\omega^2}
d\omega,
$$
where
\begin{equation}
\alpha(t)\equiv R_K+m^*(t).
\label{alphachar}
\end{equation}
Following an analogous reasoning we obtain, for the second integral on the rhs of Eq.~(\ref{charcall2}) the result
$$
I_2=2i\int_{0}^{\infty}\varphi_X(\omega,T|t)\frac{\sin\omega\alpha(T-t)}{\omega}d\omega.
$$
The substitution of these two integrals into Eq.~(\ref{charcall2}) yields our final result
\begin{eqnarray}
&&C(S,t)=S-\frac{K}{2}e^{-r(T-t)}
\nonumber\\
&&-\frac{K}{\pi}e^{-r(T-t)}
\int_{0}^{\infty}\varphi_X(\omega,T|t)\left[\cos\omega\alpha(T-t)
+\frac{\sin\omega\alpha(T-t)}{\omega}\right]\frac{d\omega}{1+\omega^2},
\label{charcallreal}
\end{eqnarray}
where $\alpha$ is given by Eq.~(\ref{alphachar}). Note that this expression only involves one real integral and it is therefore simpler and faster to compute than Eq.~(\ref{charcall2}). 

\begin{figure}[htbp]
\begin{center}
\includegraphics{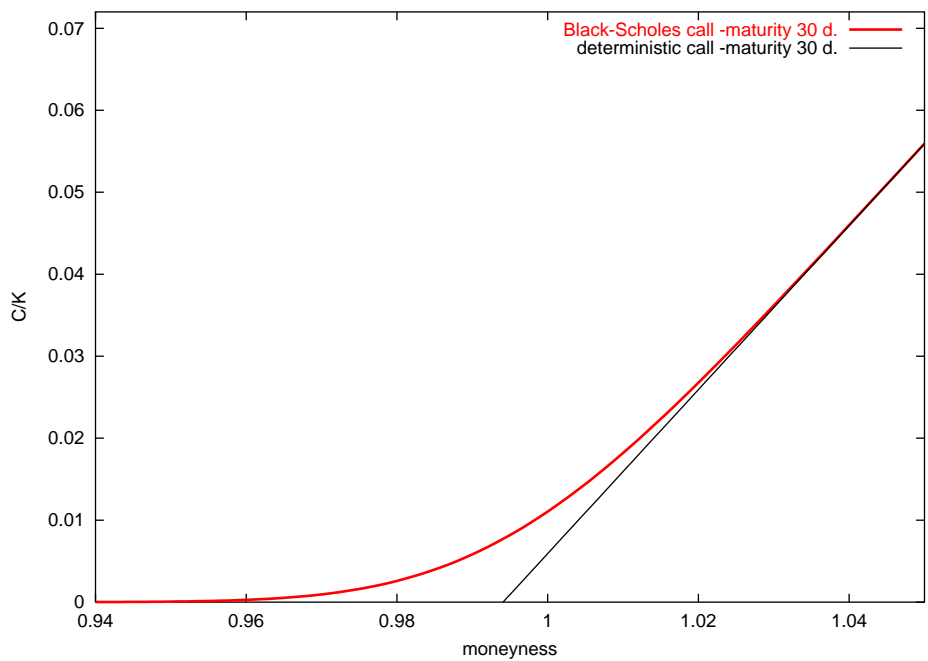}
\end{center}
\label{call}
\caption{The B-S call price in terms of the moneyness. For this graph, we take $r=5 \% \ year^{-1}$ and the volatility estimated from the S\&P-500. The B-S call price formula is given by Eq.~(\ref{BSprice}) and the deterministic price is given by Eq.~(\ref{calldeter}).}
\end{figure}

Finally, for the GBM market model, the substitution of Eqs. (\ref{fcXgbm})--(\ref{m*gbm}) and (\ref{alphachar}) into Eq. (\ref{charcallreal}) yields the Black-Scholes formula
\begin{equation}
C_{BS}(S,t)=SN(d_1)-Ke^{-rt}N(d_2),
\label{BSprice}
\end{equation}
where 
$$
N(x)=(1/\sqrt{2\pi})\int_{-\infty}^xe^{-z^2/2}dz,
$$
is the probability integral, and
\begin{equation}
d_1=\frac{rt+R_K+\sigma^2/2}{\sqrt{\sigma^2t}},
\label{d1}
\end{equation}
\begin{equation}
d_2=\frac{rt+R_K-\sigma^2/2}{\sqrt{\sigma^2t}},
\label{d2}
\end{equation}
In Fig.~\ref{call}, we plot this classic option price using the volatility estimated from Standard and Poors-500 tick by tick data (January 1986 - December 1996). Time to maturity is assumed to be 30 days and we compare the resulting call price with the deterministic call price given by Eq.~(\ref{calldeter}).

\section{Option pricing on heavy-tailed stocks\label{heav}}

We will now apply the formalism developed hitherto to pricing options whose underlying assets have probability distributions showing fat tails. In such cases, the ideal market conditions upon which the B-S theory is based are no longer valid and B-S formula is not reliable. We want to elucidate what this implies on the price of the option. 

The first market model incorporating heavy tails in a natural way was given by Mandelbroth in the early sixties \cite{mandelbrot,famalevy} who, based on Pareto-L\'evy stable laws, obtained a Leptokurtic distribution. In that model, the zero-mean return defined above has the following characteristic function
\begin{equation}
\varphi_X(\omega,t)=\exp(-kt\omega^\alpha),
\label{levy}
\end{equation}
$(1<\alpha<2, k>0)$. There is, however, a drawback in this model: no finite moments exist beyond the first one. This certainly is a severe limitation specially for the practitioner. Moreover, testing the Pareto-L\'evy distribution against data has always resulted with the same conclusion: the tails are far too heavy compared with actual data. Although, as Mantegna and Stanley have clearly shown \cite{mante95}, the L\'evy distribution fits very well the center of empirical distributions -- much better than the Gaussian distribution of the GBM. Furthermore, L\'evy distributions show a feature recently observed in data, specially in intraday data, and this is the self-scaling behavior of the probability distribution at different times~\cite{mante95,scalas,gallu}. Nevertheless, and regardless these advantages, the absence of finite moments seriously affects obtaining a price for options. Indeed, in this case the average $\langle e^{X(t)}\rangle$ does not exist (see Eqs. (\ref{expX})--(\ref{m*})) and option price~(\ref{call2}) is useless. Therefore, any attempt to price options through L\'evy processes has to be done by means of truncated L\'evy distributions \cite{matacz} with the added inconvenience that truncation is an {\it ad hoc} procedure introducing high degrees of arbitrariness. 

In a recent work we have presented a market model that explains the appearance of fat tails and self-scaling but still keeping all moments finite \cite{mmp,mmmac}. In that model we assume that all random changes in the stock price are modelled by a continuos superposition of different shot-noise sources, each source corresponding to the detailed arrival of information \cite{mmp}. The zero-mean return is given by 
\begin{equation}
X(t)=\int_{-\infty}^{n(t)}Y(u,t)du,
\label{Xmiquel}
\end{equation}
where $n(t)$ represents the maximum ``number" of noise sources taken up to time $t$ and it is an increasing function of time to be determined by the self-scaling property (see below). For any fixed time $t$, $Y(u,t)$ are independent random variables for different values of parameter $u$ and for any fixed value of $t$, $Y(u,t)$ is a white shot-noise process represented by a countable superposition of Poisson pulses of rectangular shape:
\begin{equation}
Y(u,t)=\sum_{k=1}^{\infty}A_k(u)\Theta\left[t-T_k(u)\right],
\label{Y}
\end{equation}
where $\Theta(t)$ is the Heaviside step function, $T_k(u)$ marks the onset of the $k$th pulse, and $A_k(u)$ is its amplitude. Both $T_k(u)$ and $A_k(u)$ are independent and identically
distributed random variables with probability density functions given by $h(x,u)$ and $\psi(u,t)$, respectively.

We assume that the occurrence of jumps is a Poisson process, in this case the shot-noise $Y(u,t)$ is Markovian, and the pdf for the time interval between jumps, $\psi(u,t)dt=\mbox{P}\{t<T_k(u)-T_{k-1}(u)<t+dt\}$, is exponential:
$$
\psi(u,t)=\lambda(u)e^{-\lambda(u)t}\qquad(t\geq 0),
$$
where $\lambda(u)$ is the mean jump frequency. We recall that jump amplitudes $A_k(u)$ are identically distributed (for all $k=1,2,3,\cdots$) and independent random variables (for all $k$ and $u$).
In what follows we will assume that they have zero mean and a pdf, $h(u,x)dx=\mbox{P}\{x<A_k(u)<x+dx\}$, depending on a single ``dimensional" parameter which, without loss of generality, we assume to be the standard deviation of jumps $\sigma(u)=\sqrt{\langle A_k^2(u)\rangle}$. That is,
$$
h(u,x)=\frac{1}{\sigma(u)}h\left[\frac{x}{\sigma(u)}\right].
$$
The function $n(t)$ appearing in Eq. (\ref{Xmiquel}) is determined by imposing on $X(t)$ self-scaling properties. It reads (see \cite{mmp} for details)
\begin{equation}
n(t)=\frac{1}{\alpha}\ln bt.
\label{n}
\end{equation}
Under these assumptions we have shown that the characteristic function of the zero mean return $X(t)$ is \cite{mmp} 
\begin{equation}
\varphi_X(\omega,t)=\exp\left\{-abt\int_{0}^{(bt)^{1/\alpha}}
z^{-1-\alpha}[1-\tilde{h}(\omega z)]dz\right\},
\label{final}
\end{equation}
where $\tilde{h}(\omega)$ is the characteristic function of the jump distribution $h(x)$. We thus see that the model depends on three positive constants: $a,b$, and $\alpha<2$ and an unknown distribution $h(x)$ to be guessed from data. 

Let us summarize the main results and consequences of the model. First, the model contains the L\'evy process as a special case. Indeed if $a\rightarrow 0$ and $b\rightarrow\infty$ in such a way that $ab$ is finite, Eq. (\ref{final}) yields, for any jump distribution $h(x)$, the Pareto-L\'evy distribution 
(\ref{levy}) with
$$
k=ab\int_0^{\infty}z^{-1-\alpha}[1-\tilde{h}(z)]dz.
$$
Since the limit $b\rightarrow\infty$ implies that $n(t)\rightarrow\infty$ ({\it cf.} Eq. (\ref{n})) we see from Eq. (\ref{Xmiquel}) that L\'evy process are continuos an unbounded superposition of families of white Poissonian shot noises\footnote{We use the adjective ``unbounded" in the sense that there is no maximum number of shot-noise sources, {\it i.e.,} $n(t)=\infty$ in Eq. (\ref{Xmiquel}).}.

Second, the volatility of the zero-mean return is given by 
\begin{equation}
\langle X^2(t)\rangle=\frac{a}{2-\alpha}\ (bt)^{2/\alpha}, 
\label{13}
\end{equation}
which proves that $\alpha<2$ and that the volatility shows super-diffusion. The anomalous diffusion behavior of the empirical data (at least at small time scales) was first shown by Mantegna and 
Stanley~\cite{mante96}. Third, kurtosis is constant and given by
\begin{equation}
\gamma_2=\frac{(2-\alpha)^2\tilde{h}^{(iv)}(0)}{(4-\alpha)a}.
\label{kurtosis}
\end{equation}
Thus $\gamma_2>0$ for all $t$, in other words, we have a Leptokurtic distribution ({\it i.e.,} heavy tails) in all time scales. Fourth, the return probability distribution scales as
\begin{equation}
\varphi_X(\omega,t)=\varphi_X((bt)^{1/\alpha}\omega)
\label{scale}
\end{equation} 
and the model becomes self-similar~\cite{mante95,scalas,gallu}.

As to the asymptotic behavior of our distribution. It can be shown from Eq.~(\ref{final}) that the center of the distribution, defined by $|x|<(bt)^{1/\alpha}$, is again approximated by the L\'evy distribution defined above. On the other hand the tails of the distribution are solely determined by the jump pdf $h(u)$ by means of the expression
\begin{equation}
p_X(x,t)\sim\frac{abt}{|x|^{1+\alpha}}
\int_{|x|/\sigma_m}^\infty u^\alpha
h(u)du,\qquad(|x|\gg (bt)^{1/\alpha}).
\label{14}
\end{equation}
Therefore, return distributions present fat tails and have finite moments if jump distributions behave in the same way.  This, in turn, allows us to make statistical hypothesis on the form of $h(u)$ based on the empirical form and moments of the pdf. 

\begin{figure}
\begin{center}
\includegraphics[width=12.5cm]{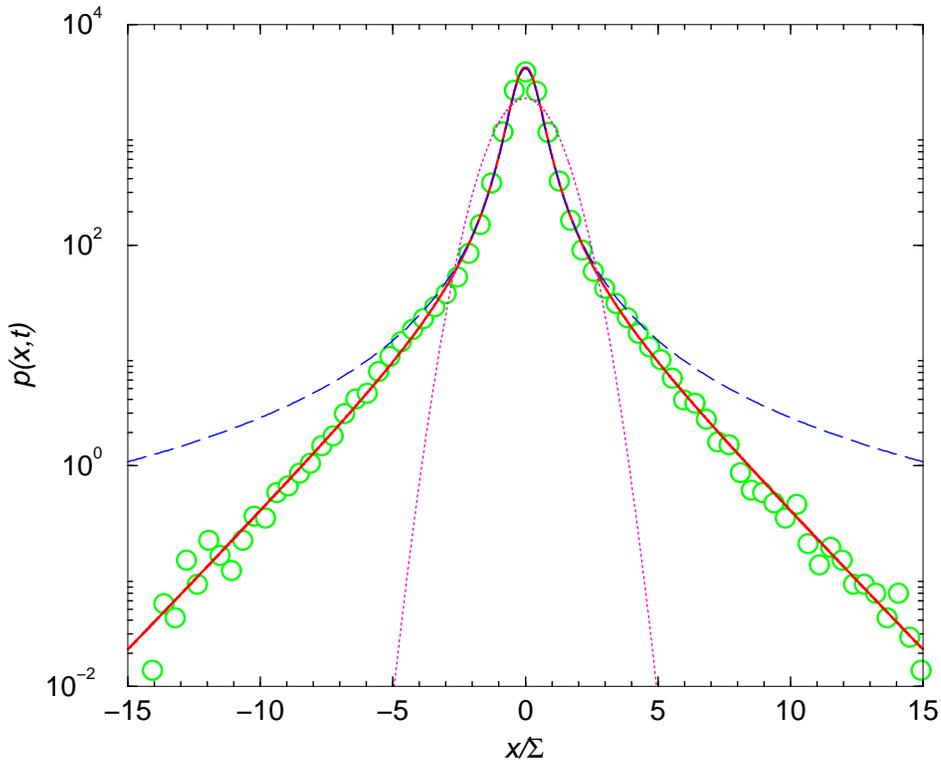}
\end{center}
\caption{Probability density function $p(x,t)$ for $t=1$ min. Circles represent empirical data from S\&P 500 cash index (January 1988 to December 1996). $\Sigma$ is the standard deviation of empirical data. Dotted line corresponds to the Guassian density. Dashed line is the L\'{e}vy distribution and the solid line is the Fourier inversion of Eq.~(\ref{final}) with a gamma distribution of jumps.}
\label{fig2}
\end{figure}

In Fig.~\ref{fig2}, we plot the probability density $p_X(x,t)$ of the  S\&P 500 cash index returns $X(t)$ observed at time $t=1$ min (circles). $\Sigma=1.87\times 10^{-4}$ is the standard deviation of the empirical data. Dotted line corresponds to a Gaussian density with standard deviation given by $\Sigma$. Solid line shows the Fourier inversion of Eq.~(\ref{final}) with $\alpha=1.30$, $\sigma_m=9.07\times 10^{-4}$, and $a=2.97\times 10^{-3}$. We use the gamma distribution of the absolute
value of jump amplitudes,
\begin{equation}
h(u)=\mu^\beta|u|^{\beta-1}e^{-\mu|u|}/2\Gamma(\beta),
\label{beta}
\end{equation}
with $\beta=2.39$, and $\mu=\sqrt{\beta(\beta+1)}=2.85$. Dashed line represents a symmetrical L\'evy stable distribution of index $\alpha=1.30$ and the scale factor $k=4.31\times 10^{-6}$~\cite{mmp}. We note that the values of $\sigma_m$ and $\Sigma$ predict that the Pareto-L\'evy distribution fails to be an accurate description of the empirical pdf for $x\gg 5\Sigma$ (see Eq.~(\ref{14})). We have chosen a gamma distribution of jumps because (i) as suggested by the empirical data analyzed, the tails of $p(x,t)$ decay exponentially, and (ii) one does not favor too small size jumps, {\it i.e.}, those jumps with almost zero amplitudes.

We will now obtain a price for a European call assuming that the underlying has a zero-mean return given by Eq. (\ref{Xmiquel}). In order to apply the formalism derived in Sect. 2 and 3 we have to see whether our $X(t)$ allows arbitrage, in other words, we have to prove that $X(t)$ has unbounded variations (see Sec. 2 and Ref. \cite{harrison2}). We proceed as follows, the variation of $X(t)$ is given by its time derivative $\dot{X}(t)$. From Eqs. (\ref{Xmiquel})--(\ref{Y}) we have
$$
\dot{X}(t)=\dot{n}(t)Y(n(t),t)+\int_{-\infty}^{n(t)}A_k(u)\delta[t-T_k(u)].
$$
Following an analogous procedure to that of Refs. \cite{mmp} and \cite{mmmac} we can see that the auto correlation of the derivative process is, in terms of the volatility $\langle X^2(t)\rangle$ ({\it cf.} Eq. (\ref{13})), given by
$$
\langle\dot{X}(t+\tau)\dot{X}(t)\rangle=
\frac{\langle X^2(t)\rangle}{t}\left[\frac{(2-\alpha)a}{\alpha t}+\delta(\tau)\right],
$$
which clearly shows that $X(t)$ is a process of unbounded variation.

Assuming a gamma distribution of jumps, Eq. (\ref{beta}), it can be proved that the characteristic function of $X(t)$ given by Eq. (\ref{final}) explicitly reads
\begin{equation}
\varphi_X(\omega, t)=\exp\left\{\frac{a}{\alpha}\left[1-
{\sf Re}\left\{F\left(\beta,-\alpha;1-\alpha;-\frac{i\omega(bt)^{1/\alpha}}{\mu}\right)\right\}\right]\right\},
\label{final2}
\end{equation}
where $F(a,b;c;z)$ is the hypergeometric function and ${\sf Re}(\cdot)$ is the real part. The substitution of this equation into Eq. (\ref{charcallreal}) results in the price of the call. Unfortunately, one cannot evaluate analytically the resulting integrals and numerical work should be performed. 

\begin{figure}
\begin{center}
\includegraphics{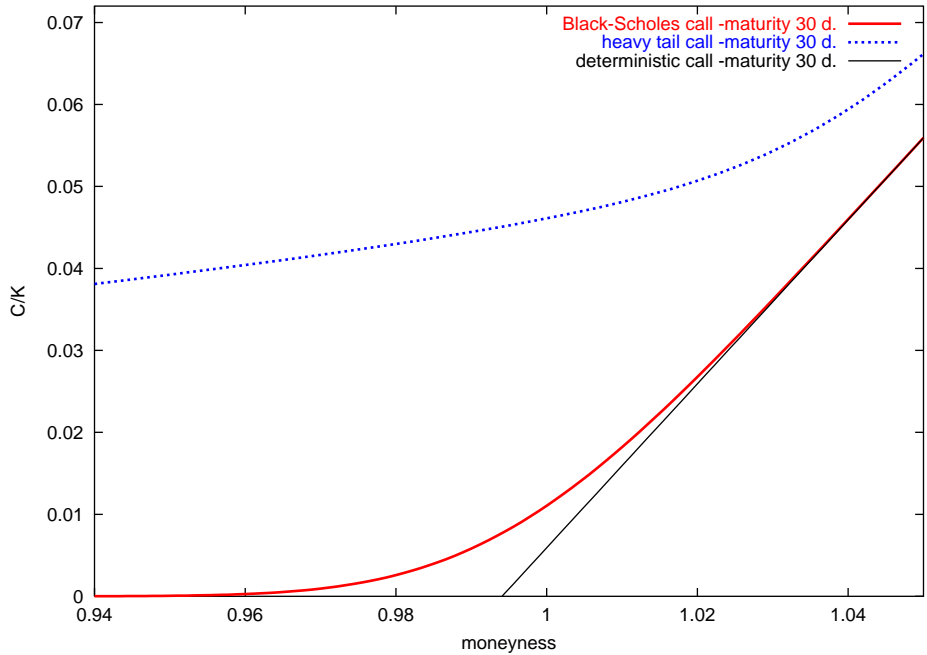}
\end{center}
\caption{The heavy tail and B-S call prices in terms of the moneyness when maturity is in 30 days. For this graph, we take $r=5 \% \ year^{-1}$ and the parameters estimated for the S\&P-500 from January 1988 to December 1996 ($\sigma=3.54\times 10^{-3} \mbox{days}^{-1/2}$).}
\label{call-heavy}
\end{figure}

\begin{figure}
\begin{center}
\includegraphics{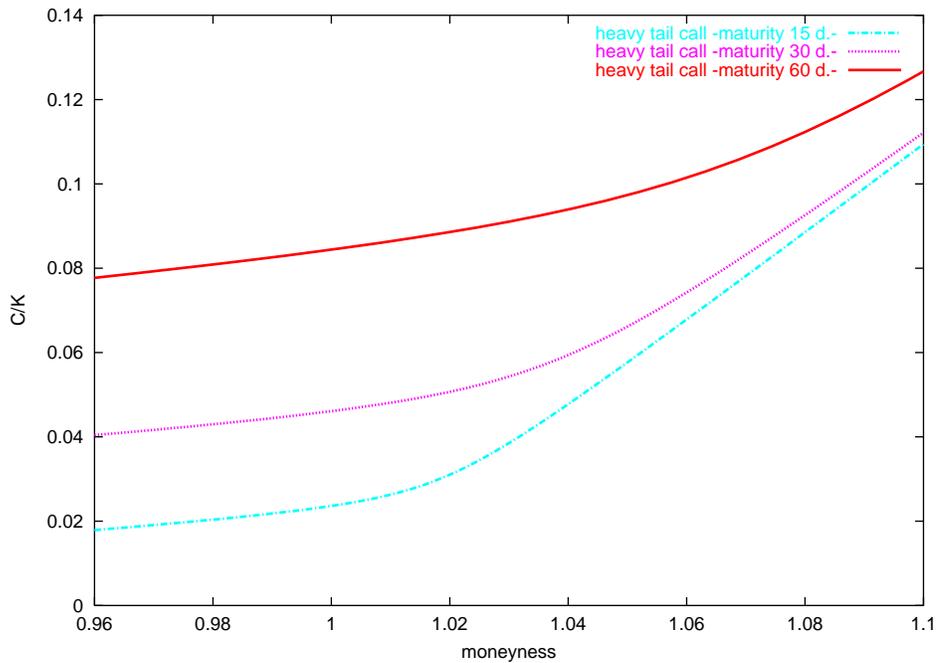}
\end{center}
\caption{The heavy tail call price in terms of the moneyness. For this graph, we take $r=5 \% \ year^{-1}$ and the parameters estimated for the S\&P-500.}
\label{calls-heavy}
\end{figure}

\begin{figure}
\begin{center}
\includegraphics{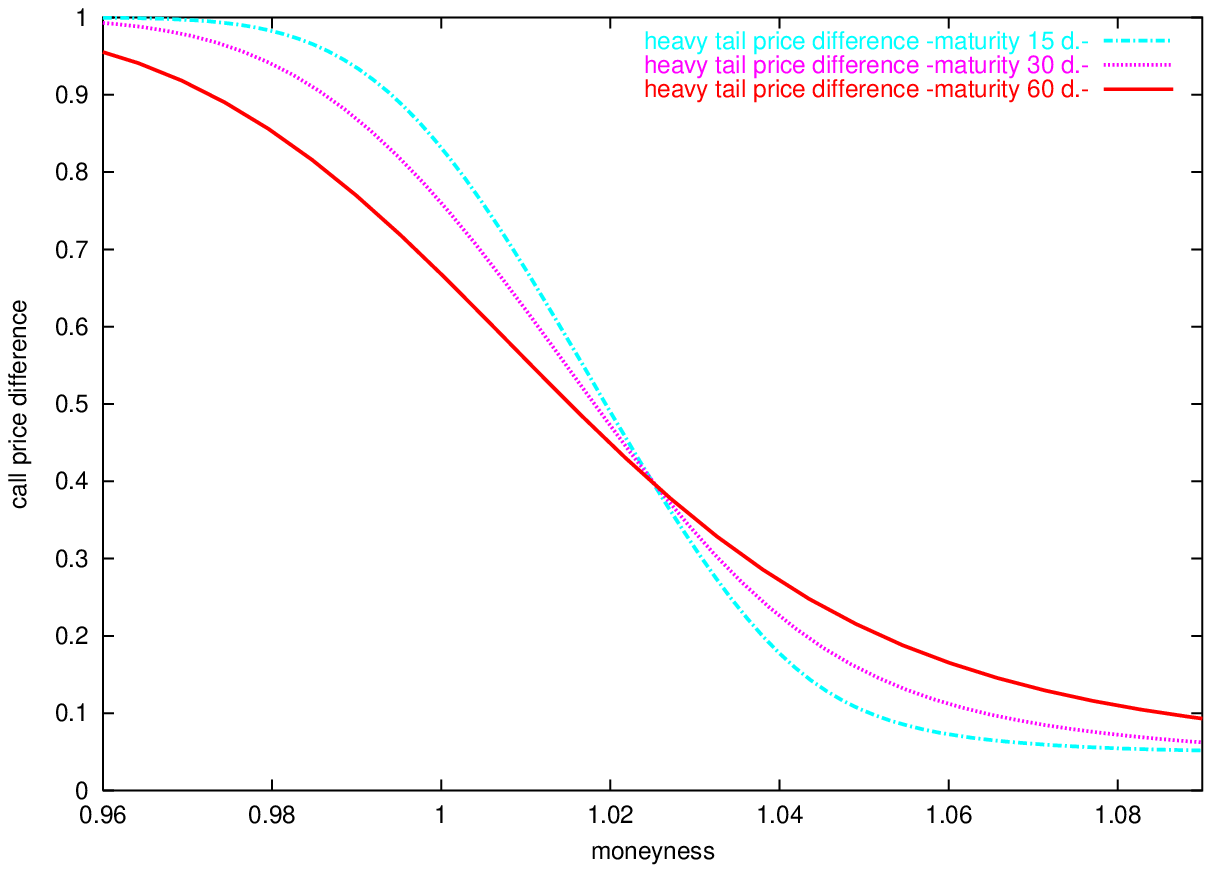}
\end{center}
\caption{The relative differences, $\Delta=(C_{Tail}-C_{BS})/C_{Tail}$,  in terms of the moneyness $S/K$ ($r=5 \% \ year^{-1}$, $\sigma=3.54\times 10^{-3} \mbox{days}^{-1/2}$).}
\label{call-diff}
\end{figure}

In Fig.~\ref{call-heavy}, we plot the call price $C_{Tail}(S,t)$ obtained by numerical integration of Eq.~(\ref{charcallreal}) with Eq.~(\ref{final2}) in terms of moneyness $S/K$. We compare this price with B-S price with the same time to maturity, $T-t$, volatility and risk-free interest rate. We see that the inclusion of heavy tails in the probability distribution substantially increases option prices. In Fig.~\ref{calls-heavy}, we plot $C_{Tail}(S,t)$ as a function of $S/K$ for three different times to maturity. We confirm that in this case (as in B-S case) an increase of time to maturity implies an increase of the call price.

Finally in Fig.~\ref{call-diff} we plot the relative difference
$$
\Delta=(C_{Tail}-C_{BS})/C_{Tail}
$$
as a function of moneyness. We observe that for OTM, ATM, and some ITM options $\Delta$ increases as $T-t$ increases, while for options well in the money $\Delta$ becomes smaller with $T-t$. There is thus a ``crossover" which, in our example, is located around $S/K=1.03$. We also note that $\Delta\rightarrow 0$ as $S/K\rightarrow\infty$ very slowly which implies the great persistence of the heavy-tail effect in the price of the option. 

\section{Option pricing on correlated stocks\label{cor}}

As we have mentioned in Section 1, the classical option formula of Black, Scholes and Merton is based on the ``efficient market hypothesis" which means that the market incorporates instantaneously any information concerning future market evolution \cite{fama}. However, as empirical evidence shows, real markets are not efficient, at least at short times \cite{fama91}. Indeed, market efficiency is closely related to the assumption of totally uncorrelated price variations (white noise). But white noise is only an idealization since, in practice, no actual random process is completely white. For this reason, white processes are convenient mathematical objects valid only when the observation time is much larger than the auto correlation time of the process. And, analogously, the efficient market hypothesis is again a convenient assumption when the observation time is much larger than time spans in which ``inefficiencies'' ({\it i.e.}, correlations, delays, etc.) occur.

Note that auto correlation in the underlying driving noise is closely related to the predictability of asset returns, of which there seems to be ample evidence \cite{breen,campbell}. Indeed, if for some particular stock the price variations are correlated during some time $\tau$, then the price at time $t_2$ will be related to the price at a previous time $t_1$ as long as the time span $t_2-t_1$ is not too long compared to the correlation time $\tau$. Hence correlation implies partial predictability.

In Ref.~\cite{corjof}, we have derived a nontrivial option price by relaxing the efficient market hypothesis and allowing for a finite, non-zero, correlation time of the underlying noise process. As a model for the evolution of the market we chose the Ornstein-Uhlenbeck (O-U) process for essentially two: (a) O-U
noise is still a Gaussian random process with an arbitrary correlation time $\tau $ and it has the property that when $\tau=0$ the process becomes Gaussian white noise, as in the original Black-Scholes option case. (b) The O-U process is, by virtue of Doob's theorem, the only Gaussian random process which is simultaneously Markovian and stationary. In this sense the O-U process is the simplest generalization of Gaussian white noise. 

We thus assume that the underlying asset price is not driven by Gaussian white noise $\xi(t)$ but by O-U noise $V(t)$. In other words, instead of Eq.~(\ref{gbmR}), we now have the following singular two-dimensional diffusion process:
\begin{equation}
\dot{R}(t)=\mu+V(t),
\label{OUR}
\end{equation}
\begin{equation}
\dot{V}(t)=\frac{1}{\tau}\left[-V(t)+\sigma\xi(t)\right],
\label{OUV}
\end{equation}
where $\tau\geq 0$ is the correlation time. More precisely, $V(t)$ is O-U noise in the stationary regime, which is a Gaussian colored noise with zero mean and correlation function:
\begin{equation}
\langle V(t_1)V(t_2)\rangle=\frac{\sigma^2}{2\tau}e^{-|t_1-t_2|/\tau}. 
\label{correlv}
\end{equation}
Note that, when $\tau=0$, this correlation goes to $\sigma^2\delta(t_1-t_2)$ and we thus recover the one-dimensional diffusion discussed above. Therefore, the case of positive $\tau$ is a measure of the inefficiencies of the market. The combination of Eqs.~(\ref{OUR}) and (\ref{OUV}) leads to a second-order stochastic differential equation for $R(t)$ 
\begin{equation}
\tau\ddot{R}(t)+\dot{R}(t)=\mu+\sigma\xi(t).
\label{ddotR}
\end{equation}
From this equation, we clearly see that when $\tau=0$ we recover the one-dimensional diffusion case~(\ref{gbmR}). In the opposite case when $\tau=\infty$, Eq.~(\ref{OUV}) shows that $\dot{V}(t)=0$. Thus, $V(t)$ is constant which we may equal it to zero. Hence, $R(t)=\mu t$ and the underlying price, $S(t)=S_0e^{\mu t}$, evolves as a riskless security. We also observe that the O-U process $V(t)$ is the random part of the return velocity, and sometimes we refer to $V(t)$ as the ``velocity'' of the return process $R(t)$.

One may argue that the O-U process~(\ref{OUR})--(\ref{OUV}) is an inadequate asset model since the return $R(t)$ given by Eq.~(\ref{OUR}) is a continuous random process with bounded variations. As we have explained in Section 2, continuous processes with bounded variations allow arbitrage opportunities and this is an undesirable feature for obtaining a fair price. Thus, for instance, in this present case  arbitrage would be possible within a portfolio containing bonds and stock whose strategy at time $t$ is buying (or selling) stock shares when $\mu+V(t)$ is greater (or lower) than the risk-free bond rate~\cite{harrison2}. However, the problem is that in the practice the return velocity $V(t)$ is non tradable and its evolution is ignored. In other words, in real markets the observed asset dynamics does not show any trace of the velocity variable. Indeed, knowing $V(t)$ would imply knowing the value of the return $R(t)$ at two different times, since
$$
V(t)=\lim_{\epsilon\rightarrow 0^+}\frac{R(t)-R(t-\epsilon)}{\epsilon}-\mu.
$$
Obviously, this operation is not performed by traders who only manage portfolios at time $t$ based on prices at $t$ and not at any earlier time. This feature allows us to perform a projection of the two-dimensional diffusion process $[R(t),V(t)]$ onto a one-dimensional equivalent process $\bar{R}(t)$ independent of the velocity $V(t)$. We have shown elsewhere \cite{corjof} the projected process $\bar{R}(t)$ is equal to the actual return $R(t)$ in mean square sense, $\langle[R(t)-\bar{R}(t)]^2\rangle=0$ for any time $t$, and it obeys the following one-dimensional SDE\footnote{Since $R(t)$ and $\bar{R}(t)$ are equal in mean square sense we will drop the bar on $\bar{R}$ as long as there is no confusion. Thus, we will use $R$ for the projected process as well.}
\begin{equation}
\dot{R}(t)=\mu+\sqrt{\dot{\kappa}(T-t)}\xi(t),
\label{sde}
\end{equation}
where
\begin{equation}
\dot{\kappa}(t)=\sigma^2\left(1-e^{-t/\tau}\right).
\label{dotkappa}
\end{equation}  
In this way, we have projected the two-dimensional O-U process $(S,V)$ onto a one-dimensional price process which is a Wiener process with time varying volatility. Therefore, the return given by Eq.~(\ref{sde}) is driven by a noise of unbounded variation, the Wiener process, and the O-U projected process is still a suitable starting point for option pricing since it does not permit arbitrage. We also note that we need to specify the final condition of the process because the volatility $\sqrt{\dot{\kappa}}$ is a function of the time to maturity $T-t$, and this implies that the projected asset model depends on each particular contract.

The zero-mean return (\ref{X}) associated with the projected process (\ref{sde}) follows the SDE
$$
\dot{X}(t)=\sqrt{\dot{\kappa}(T-t)}\xi(t),
$$
and its density, $p_X(x,T|t)$, satisfies the backward equation 
$$
\frac{\partial p_X}{\partial t}=
-\frac 12 \dot{\kappa}(T-t)\frac{\partial^2p_X}{\partial x^2},
$$
with final condition $p_X(x,T|T)=\delta(x)$. The solution to this problem 
reads
\begin{equation}
p_X(x,T|t)=\frac{1}{\sqrt{2\pi\kappa(T-t)}}
\exp\left[-\frac{x^2}{2\kappa(T-t)}\right],
\label{pdfXcorr}
\end{equation}
where ({\it cf.} Eq. (\ref{dotkappa}))
\begin{equation}
\kappa(t)=\sigma^2\left[t-\tau\left(1-e^{t/\tau}\right)\right].
\label{kappa}
\end{equation}
The corresponding characteristic function reads
$$
\varphi_X(\omega,T|t)=\exp[-\kappa(T-t)\omega^2/2],
$$
and from Eq. (\ref{m*}) we get
\begin{equation}
m^*(T-t)=r(T-t)-\kappa(T-t)/2.
\label{m*corr}
\end{equation}

In this case the price of the European option is obtained substituting Eqs. (\ref{pdfXcorr}) and 
(\ref{m*corr}) into Eq. (\ref{call2}). We have
\begin{equation}
C_{OU}(S,t)=S\ N(d^{OU}_1)-Ke^{-r (T-t)}\ N(d^{OU}_2),
\label{call4} 
\end{equation}
where $N(z)$ is the probability integral, and
\begin{equation}
d^{OU}_1=\frac{\ln(S/K)+r (T-t)+\kappa(T-t)/2}{\sqrt{{\kappa}(T-t)}},
\label{d1OU}
\end{equation}
\begin{equation}
d^{OU}_2=d^{OU}_1-\sqrt{{\kappa}(T-t)},
\label{d2OU} 
\end{equation}
with $\kappa (t)$ given by Eq.~(\ref{kappa}).

When $\tau=0$, the variance becomes $\kappa(t)=\sigma^2 t$ and the price in Eq. (\ref{call4}) reduces to the Black-Scholes price given by Eq. (\ref{BSprice}). We observe that the O-U price in Eq. (\ref{call4}) has the same functional form as B-S price in Eq. (\ref{BSprice}) when $\sigma^2 t$ is replaced by $\kappa(t)$. 

\begin{figure}
\begin{center}
\includegraphics{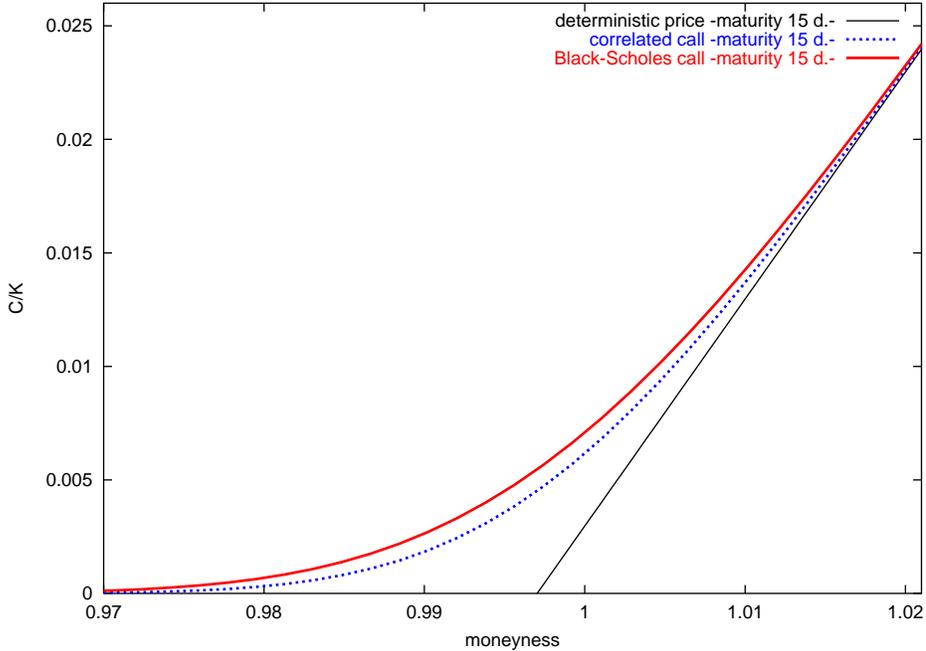}
\end{center}
\caption{The correlated and B-S call prices in terms of the moneyness when maturity is in 15 days and correlation time is 2 days. For this graph, we take $r=5 \% \ year^{-1}$ and the parameters estimated for the S\&P-500 ($\sigma=3.54\times 10^{-3} \mbox{days}^{-1/2}$).}
\label{call-correlated}
\end{figure}

In the opposite case, $\tau=\infty$, where there is no random noise but a deterministic and constant driving force (in our case it is zero), Eq. (\ref{call4}) reduces to the deterministic price ({\it cf.} Eqs. (\ref{payoff}) and~(\ref{call1})) 
\begin{equation}
C_d(S,t)=\max[S-Ke^{-r(T-t)},0].
\label{calldeter}
\end{equation}
We have proved elsewhere \cite{corjof} that the O-U price is an intermediate price between B-S price and the deterministic price (see Fig.~\ref{call-correlated})
\begin{equation}
C_{d}(S,t)\leq C_{OU}(S,t)\leq C_{BS}(S,t),
\label{bounds} 
\end{equation}
for all $S$ and $0\leq t\leq T$.

\begin{figure}
\begin{center}
\includegraphics{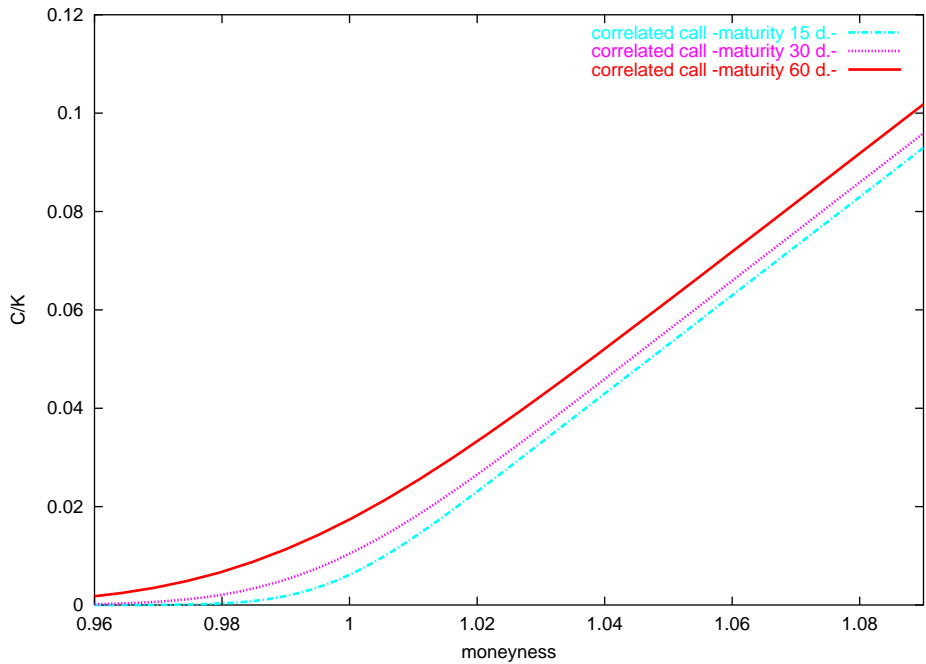}
\end{center}
\caption{The correlated call prices in terms of the moneyness with correlation time of 2 days. For this graph, we take $r=5 \% \ year^{-1}$ and the parameters estimated for the S\&P-500 ($\sigma=3.54\times 10^{-3} \mbox{days}^{-1/2}$).}
\label{calls-correlated}
\end{figure}

\begin{figure}
\begin{center}
\includegraphics{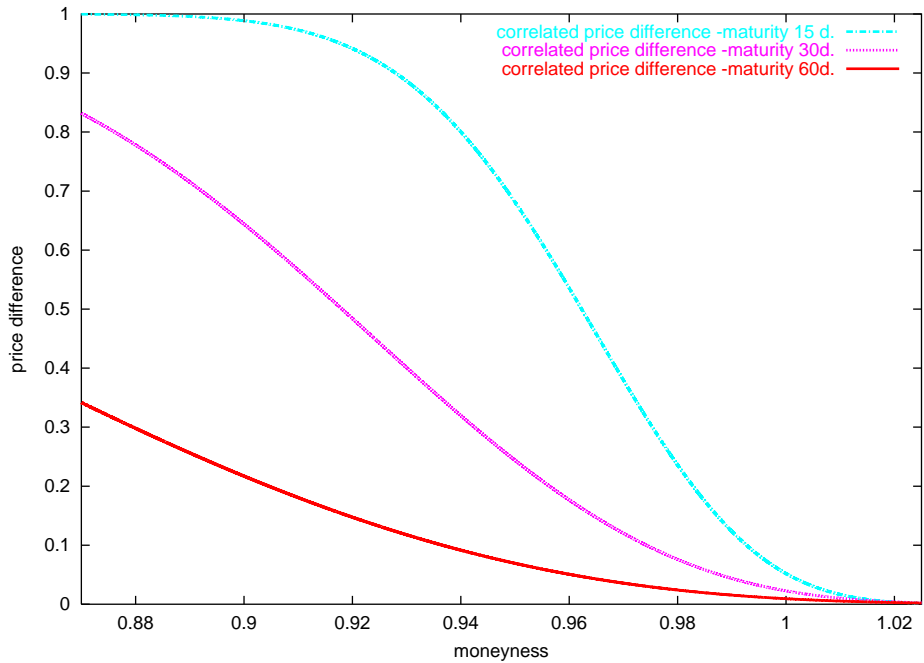}
\end{center}
\caption{Relative differences, $D=(C_{BS}-C_{OU})/C_{BS}$, in terms of moneyness ($\tau=2$ days, $r=5 \% \ year^{-1}$, and $\sigma=3.54\times 10^{-3} \mbox{days}^{-1/2}$).}
\label{corr-diff}
\end{figure}

In Fig. \ref{calls-correlated}, we plot $C_{OU}$ in terms of $S/K$ for three different values of the time to maturity. We see in the figure that, analogously to B-S and heavy tail cases, the O-U price increases with maturity. In any case, {\it the assumption of uncorrelated underlying assets (B-S case) overprices any call option}. This confirms the intuition understanding that correlation implies more predictability and therefore less risk and, finally, a lower price for the option. 

In fact, one can quantify this overprice by evaluating the relative difference
$$
D=(C_{BS}-C_{OU})/C_{BS}.
$$
Figure~\ref{corr-diff} shows the ratio $D(S,t)$, for a fixed time to expiration, plotted as a function of the moneyness, $S/K$. Note that the relative difference $D$ increases as $T-t$ decreases. Moreover, contrary to the fat tail case, $D$ quickly fades away with moneyness and it is negligible for ITM options showing no crossover. 

\section{Conclusions\label{conc}}

We have performed an analytical study of option pricing on a more realistic setting than that of Black and Scholes theory. Our starting point has been the martingale option price conveniently modified to become more practical to implement by means of Fourier analysis. We have applied these analytical tools to study the effect on option price of long tails in the underlying probability distribution. For this, we have used a market model we recently developed \cite{mmp} which combines fat tails, finite moments and self-similarity. The second problem we wanted to address is the effect of colored noise on option pricing, since there seems to be empirical evidence of mild auto correlation in stocks \cite{lo}. We have thus rederived, using the methods herein presented, our previous result on option pricing on stocks driven by O-U noise. 

One overall conclusion is that, taking the Black-Scholes price as a benchmark, heavy tails overprice by far options, while colored driving noise underprices it. The scheme is
$$
S\geq C_{Tail}\geq C_{BS}\geq C_{OU}\geq C_d,
$$
where $S$ is the underlying price when the option is bought and $C_d$ is the deterministic price given by Eq. (\ref{calldeter}). Moreover, the relative difference with B-S price is in both cases more important for OTM and ATM options than for ITM options. In any case, there is a great persistence of fat-tail effects on the price of the options. 

There are some open questions, one of which being the extension of the above formalism to other option classes (American, Asian, etc.). Another point is to investigate whether fat tails and correlations produce the ``smile effect" in the implied volatility. We believe that this is more likely to happen in the case of heavy-tails. This point is under present investigation.
 
\begin{ack}

The authors acknowledge helpful discussions with Miquel Montero on the results herein derived for the market model showing fat tails. This work has been supported in part by Direcci\'on General de Investigaci\'on under contract No. BFM2000-0795, by Generalitat de Catalunya under contract No. 2000 SGR-00023 and by Societat Catalana de F\'{\i}sica (Institut d'Estudis Catalans). 

\end{ack}

\appendix

\section{Derivation of Eq.~(\ref{I})}

Let us derive Eq. (\ref{I}). We start from Eq. (\ref{I0})
$$
I=\frac{K}{2\pi}\int_{-\infty}^{\infty}\varphi^*(\omega,T|t)e^{i\omega R_K}d\omega
\int_{0}^{\infty}e^{i\omega z}\Bigl(e^{-z}-1\Bigr)dz,
$$
but
$$
\int_{0}^{\infty}e^{i\omega z}\Bigl(e^{-z}-1\Bigr)dz=\frac{1}{1-i\omega}-
\int_{0}^{\infty}e^{i\omega z}dz,
$$
and
$$
\int_{0}^{\infty}e^{i\omega z}dz=\lim_{\varepsilon \rightarrow 0^+}
\int_{0}^{\infty}e^{-(\varepsilon-i\omega)z}dz=
\lim_{\varepsilon \rightarrow 0^+}\frac{i}{\omega+i\varepsilon}=\frac{i}{\omega+i0},
$$
where this result has to be understood in the sense of generalized functions as \cite{vladimirov}
$$
\frac{1}{\omega\pm i0}=\mp i \pi \delta(\omega)+ P\left[\frac{1}{\omega}\right],
$$
where $\delta(\omega)$ is the Dirac delta function, and $P [1/\omega]$ is the Cauchy principal value, {\it i.e.}, 
$$
P [1/\omega]= 1/\omega \qquad \mbox{for }\omega\neq 0
$$
and
$$
P \left[\int_{-\infty}^{\infty} \frac{\phi(\omega)}{\omega}d\omega\right]=\lim_{\varepsilon\rightarrow 0^+} \left(\int_{-\infty}^{-\varepsilon} \frac{\phi(\omega)}{\omega}d\omega+\int_{\varepsilon}^{\infty} \frac{\phi(\omega)}{\omega} d\omega\right),
$$
where $\phi(\omega)$ is any regular function fast decaying at infinity. Therefore,
$$
\int_{0}^{\infty}e^{i\omega z}dR=\pi\delta(\omega)+i P \left[\frac{1}{\omega}\right].
$$
Hence
$$
I=-\frac{K}{2}+
\frac{K}{2\pi}\int_{-\infty}^{\infty}\varphi^*(\omega,T|t)\frac{e^{i\omega R_K}d\omega}{1-i\omega}
-i P \left[\int_{-\infty}^{\infty}e^{i\omega R_K}\frac{\varphi^*(\omega,T|t)}{\omega}d\omega\right],
$$
and using the following alternative expression for the Cauchy principal value \cite{vladimirov}
$$
P \left[\int_{-\infty}^{\infty} \frac{\phi(\omega)}{\omega}d\omega\right]=
\int_{-\infty}^{\infty} \frac{\phi(\omega)-\phi(0)}{\omega}d\omega,
$$
we obtain Eq. (\ref{I}):
$$
I=-\frac{K}{2}+
\frac{K}{2\pi}\int_{-\infty}^{\infty}\varphi^*(\omega,T|t)\frac{e^{i\omega R_K}d\omega}{1-i\omega}
-i\int_{-\infty}^{\infty}\Bigl[e^{i\omega R_K}\varphi^*(\omega,T|t)-1\Bigr]
\frac{d\omega}{\omega}.
$$

\end{document}